\documentclass[twocolumn,showpacs,aps,prl]{revtex4}
\usepackage{amsmath}
\usepackage{amssymb}
\usepackage{graphicx}
\usepackage{graphics}

\begin{document}

\title{Skeleton expansions for directed polymers in disordered media}
\author{Semjon Stepanow}
\affiliation{Martin-Luther-Universit\"{a}t Halle-Wittenberg, Institut f\"{u}r Physik,
D-06099 Halle, Germany}
\date{\today}

\begin{abstract}
Partial summations of perturbation expansions of the directed polymer in
disordered media (DPRM) enables one to represent the latter as skeleton
expansions in powers of the effective coupling constant $\Delta (t)$, which
corresponds to the binding state between two replicas in the replica field theory of DPRM,
and is equivalent to the binding state of a quantum particle in an external $\delta $%
-potential. The strong coupling phase is characterized by the exponential
dependence of $\Delta (t)$ on $t$, $\Delta (t)\sim \exp (p_{c}t)$ with $%
p_{c} $ being the binding energy of the particle. For dimensions $d>2$ the
strong coupling phase exists for $\Delta _{0}>\Delta _{c}(d)$. We compute
explicitly the mean-square displacement and the 2nd cumulant of the free
energy to the lowest order in powers of effective coupling in $d=1$. We
argue that the elimination of the terms $\exp (p_{c}t)$ in skeleton
expansions demands an additional partial summation of skeleton series.
\end{abstract}

\maketitle


The behavior of a directed polymer in a disordered medium is a fundamental
problem in statistical physics, which is related to different problems such
as non equilibrium dynamics of surfaces, growth processes such as Eden
model, growth by ballistic deposition, randomly stirred fluids (Burgers
equation), dissipative transport in the driven-diffusion equation, and the
behavior of flux lines in superconductors \cite{nelson77}-\cite{krug97}.
Despite enormous interest over the recent years (for reviews see \cite%
{hh-zhang}-\cite{krug97}) there is no basic understanding on the analytical
description of the strong coupling behavior of the directed polymer.
Moreover there is even a controversy on the existence of the upper critical
dimension $d_{c}=4$ \cite{lassig98}-\cite{fogedby05} or infinity \cite%
{tang92}-\cite{marinari02}. Very recently a connection between DPRM and
random matrices was established in \cite{spohn06}. It was also found that DPRM
possesses a multicritical behavior at the transition in high dimensions \cite%
{mukherji96}-\cite{monthus07}.

The distribution function of the free polymer end $W(\mathbf{x},t)$ the
equation
\begin{equation}
{\frac{\partial W(\mathbf{x},t)}{\partial t}}=\gamma \boldsymbol{\nabla }%
^{2}W+\mu (\mathbf{x},t)W,  \label{dp1}
\end{equation}%
where the random potential $\mu (\mathbf{x},t)$ is Gaussian distributed and
possesses the moments%
\begin{equation*}
\left[ \mu (\mathbf{x},t)\right] \,=0,\ \left[ \mu (\mathbf{x},t)\mu (%
\mathbf{x^{\prime }},t^{\prime })\right] =\Delta _{0}\delta ^{d}(\mathbf{x}-%
\mathbf{x}^{\prime })\delta (t-t^{\prime }).
\end{equation*}

The Cole-Hopf transformation $W(\mathbf{x},t)=\exp \left( \frac{\lambda }{%
2\gamma }h(\mathbf{x},t)\right) $ maps (\ref{dp1}) to the
Kardar-Parisi-Zhang equation \cite{kpz}
\begin{equation}
\frac{\partial h}{\partial t}=\gamma \boldsymbol{\nabla }^{2}h+\frac{\lambda
}{2}{(}\boldsymbol{\nabla }h{)}^{2}+\eta (\mathbf{x},t),  \label{dp2}
\end{equation}
where $h(\mathbf{x},t)$ is a single-valued function, which describes the
height profile above a basal $d$-dimensional substrate $\mathbf{x}$ in the
co-moving coordinate system, $\lambda $ is responsible for the lateral
growth, $\gamma $ is the surface tension, and the noise $\eta (\mathbf{x}%
,t)=(2\gamma /\lambda )\mu (\mathbf{x},t)$ with the correlator $\langle \eta
(\mathbf{x},t)\eta (\mathbf{x^{\prime }},t^{\prime })\rangle =2D\delta ^{d}(%
\mathbf{x}-\mathbf{x}^{\prime })\delta (t-t^{\prime })$ and $D=2\Delta
_{0}\gamma ^{2}/\lambda ^{2}$. The introduction of the new variable $\mathbf{%
v}=\boldsymbol{\nabla }h$ results in the Burgers equation \cite%
{huse-henley-fisher}
\begin{equation}
\frac{\partial v_k}{\partial t}=\gamma \boldsymbol{\nabla }^{2}v_k+\lambda
v_i\nabla_k v_i+f_k(\mathbf{x},t)  \label{dp3}
\end{equation}%
with the stirring force $\mathbf{f}(\mathbf{x},t)=\boldsymbol{\nabla }\eta (%
\mathbf{x},t)$.

The behavior of the directed polymer is determined by two competing factors:
\textit{i}) Energy win from low valued sites of the random potential, which
requires a transversal wandering of the polymer and \textit{ii}) the lost of
elastic energy due to polymer stretching. The critical exponents $\zeta $
and $\omega $, which describe the behavior of the transversal displacement
of the free polymer end and the fluctuation of the free energy
\begin{equation}
\left[ <x^{2}(t)>\right] \sim t^{2\zeta },\ \ \delta F(t)\sim t^{\omega }
\label{dp5}
\end{equation}%
are exactly known for $d=1$: $\zeta =2/3$, $\omega =1/3$ \cite%
{huse-henley-fisher}. The exponents fulfil the relation $2\zeta =\omega +1$,
which is due to the Galilean invariance.

The motivation for the present approach is the following. The
renormalization group (RG) method permits to reorganize the bare perturbation series as
expansions in powers of the effective coupling constant. In
the case of DPRM it was shown that the one-loop RG is exact \cite{somendra91}- \cite{lassig95}, but the effective coupling constant increases under
renormalization, and corresponds to the so-called run-away situation. The
effective coupling constant possesses a pole at a finite length, which restricts the renormalization until this length. In this paper
instead of using RG we perform partial summations of perturbation
expansions of DPRM and arrive at the skeleton expansions in powers of the
effective coupling constant $\Delta (t)$, which corresponds to the exact
four-vertex of the replica field theory of DPRM
The
effective coupling constant $\Delta (t)$ is well-defined for all $t$, and
increases exponentially with $t$, $\Delta (t)\sim \exp (p_{c}t)$,
with $p_{c}$ being the binding energy in the $\delta $-potential. The strong
coupling phase manifests itself in the exponential increase of $\Delta (t)$
with $t$. For dimensions $d>2$ the strong coupling phase exists for $\Delta
_{0}>\Delta _{c}(d)$. We compute explicitly the mean-square displacement and
the 2nd cumulant of the free energy to the lowest order in effective
coupling in $d=1$. A fundamental problem of elimination of the exponential
terms $\exp (p_{c}t)$ from skeleton expansions remains, however, unsolved.
Relying on the problem of the perturbational study of the quantum particle
in an external $\delta $-potential we argue that the compensation of the
exponential terms demands an additional partial summation of skeleton
series, which physically corresponds to the redefinition of the ground state.

The free energy and its fluctuations can be obtained directly from the
replica partition function $\left[ Z^{n}\right] $, which is the partition
function of $n$ quantum particles with Hamiltonian%
\begin{equation*}
H_{n}=-\frac{\gamma }{2}\sum\limits_{a=1}^{n}\nabla _{a}^{2}-\frac{\Delta }{2%
}\sum\limits_{a,b=1}^{n}\delta ^{(d)}(x_{a}-x_{b}).
\end{equation*}%
The perturbation series of $\left[ Z^{n}\right] $ in powers of $\Delta _{0}$
can be represented by graphs examples of which are shown in Fig. \ref%
{pdm_Z^n}.
\begin{figure}[tbph]
\begin{center}
\includegraphics[scale=0.5]{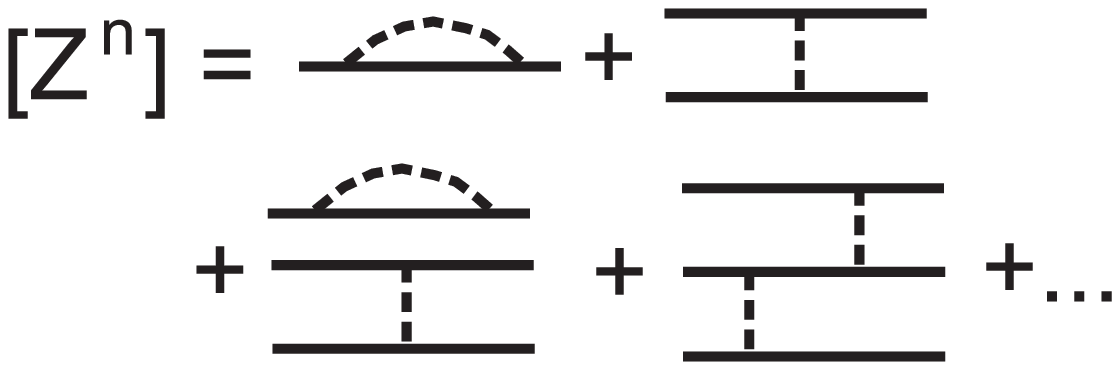}
\end{center}
\caption{Examples of graphs associated with the perturbation series of the
replica partition function.}
\label{pdm_Z^n}
\end{figure}
The transformation
\begin{equation*}
\left[ Z^{n}\right] =\left[ e^{n\ln Z}\right] =\left[ e^{-nF/k_{B}T}\right]
=\int dFp(F)e^{-nF/k_{B}T}
\end{equation*}%
shows that the inverse Laplace transform of the replica partition function $%
\left[ Z^{n}\right] $ yields the distribution function of the \ free energy $%
p(F)$ \cite{zhang90}. The cumulants $C_{k}$ of $\left[ Z^{n}\right] $, which
are defined by the equation
\begin{equation*}
\left[ Z^{n}\right] =\exp \left( \sum\limits_{k=1}^{\infty }\frac{C_{k}}{k!}%
n^{k}\right)
\end{equation*}%
read%
\begin{eqnarray*}
C_{1} &=&\left[ \ln Z\right] , \,\,
C_{2} =\left[ \ln ^{2}Z\right] -\left[ \ln Z\right] ^{2}, \\
C_{3} &=&\left[ \ln ^{3}Z\right] -3\left[ \ln ^{2}Z\right] \left[ \ln Z%
\right] +2\left[ \ln Z\right] ^{3}, \\
&&\cdots.
\end{eqnarray*}%
Thus, the free energy and its fluctuations (in units of $k_{B}T$) are given
by cumulants of the replica partition function. \

The transversal displacement of the free polymer end with one end fixed at
the origin is computed to the lowest order in $\Delta _{0}$ as%
\begin{equation*}
\left[ <x^{2}(t)>\right] =d\gamma t(1+\frac{1}{4-d}\frac{\Delta _{0}}{(4\pi
\gamma )^{d/2}}t^{1-d/2}+\cdots).
\end{equation*}%
The computation of the first two cumulants for polymer with fixed one end to
the lowest order yields
\begin{eqnarray*}
-C_{1} &=&t\frac{\Delta _{0}}{2}\delta (0)+\frac{\Delta _{0}}{(4\pi \gamma
)^{d/2}}t^{1-d/2}+\cdots, \\
C_{2} &=&\frac{\Delta _{0}}{(2\pi \gamma )^{d/2}}\frac{1}{1-d/2}%
t^{1-d/2}-\cdots.
\end{eqnarray*}%
A direct computation shows that for a polymer with both free ends the
disorder is irrelevant.

\begin{figure}[tbph]
\begin{center}
\includegraphics[scale=0.4]{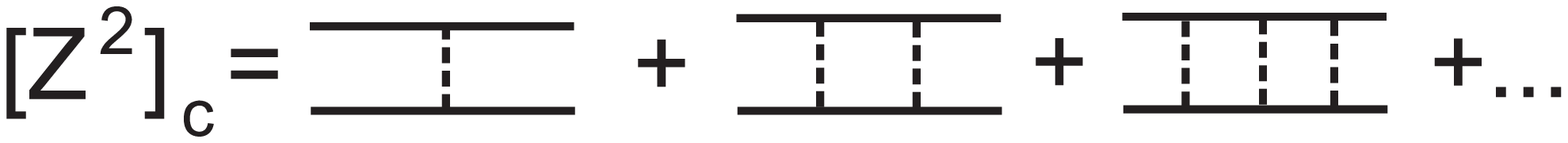}
\end{center}
\caption{The perturbation series of the four vertex in powers of $\Delta _{0}
$.}
\label{pdm_Z^2_c}
\end{figure}

Let us now consider the connected part of the four vertex $\left[ Z^{2}%
\right] _{c}$ which is shown in Fig. \ref{pdm_Z^2_c}. The graphs describe
the interaction between two polymers (quantum particles for imaginary
times). As in the two particle problem the center of mass and relative
coordinates can be separated by using the appropriate
transformation of coordinates. The part of $\left[ Z^{2}\right] _{c}$ depending on relative
coordinates coincides with the perturbation series of the Greens function of
a polymer in an external $\delta $-potential.
This series determines the renormalization of the bare coupling constant $%
\Delta _{0}$. The straightforward summation of the perturbation series in $%
d<2$ gives
\begin{equation}
\Delta (p)=\frac{\Delta _{0}}{1-\Delta _{0}(4\pi \gamma )^{-d/2}\Gamma
(1-d/2)p^{-\left( 1-d/2\right) }},  \label{dp6}
\end{equation}%
where $\Gamma (1-d/2)p^{-\left( 1-d/2\right) }$ is the Laplace transform of
the probability $G_{0}(0,t;0)\simeq t^{-d/2}$. The pole of (\ref{dp6}%
) determines the binding energy $p_{c}$ of the quantum particle in $\delta $%
-potential. The inverse Laplace transform of (\ref{dp6}) is well-defined for
all times $t$. For large $t$ in $d=1$ we obtain $\Delta (t)$ as%
\begin{equation}
\Delta (t)=\frac{\text{$\Delta $}_{0}^{3}}{2\gamma }e^{t\text{$\Delta $}%
_{0}^{2}/4\gamma }+\sqrt{\frac{\gamma }{\pi }}t^{-3/2}+\cdots  \label{dp8}
\end{equation}%
Eq. (\ref{dp6}), as it is, applies for $d<2$ only. To extend the
applicability of Eq. (\ref{dp6} to dimensions $d\geq 2$ we replace the
return probability $t^{-d/2}$ by $t^{-d/2}\exp \left( -\lambda ^{2}/t\right)
$, where $\lambda $ is a microscopic cutoff. The effective coupling constant
is given in this case by the expression%
\begin{equation}
\Delta (p)=\frac{\text{$\Delta $}_{0}}{1-2\text{$\Delta _{0}$}(4\pi \gamma
)^{-d/2}\left( \frac{p}{\lambda ^{2}}\right) ^{\frac{d-2}{4}}K_{d/2-1}\left(
2\lambda \sqrt{p}\right) },  \label{dp6d}
\end{equation}%
where $K_{n}\left( x\right) $ is the modified Bessel function of the second
kind. For $d<2$ Eq. (\ref{dp6d}) coincides with Eq. (\ref{dp6} in the limit $%
\lambda \rightarrow 0$. The necessity of the cutoff in $d=2$ follows from
the connection of DPRM to quantum particle in a weak potential well, where
in $d=2$ the binding energy depends both on the depth and the width of the
potential well, and not as in $d\leq 2$ only on the product of depth and width.
The analysis of \ the eigenvalue equation for $d>2$ (the denominator of Eq. (%
\ref{dp6d})) yields that a binding state exists in dimensions, if $\Delta
_{0}$ exceeds a threshold value $\Delta _{0}^{c}(d)$. Consequently, in
accordance with Eq. (\ref{dp8}) the effective coupling constant increases
exponentially with time for all dimensions, if $\Delta _{0}$ exceeds the
threshold value $\Delta _{0}^{c}(d)$. The latter suggests that the strong
coupling phase exists for $d>2$, if the condition $\Delta _{0}>\Delta
_{0}^{c}(d)$ applies. \

\begin{figure}[tbph]
\begin{center}
\includegraphics[scale=0.4]{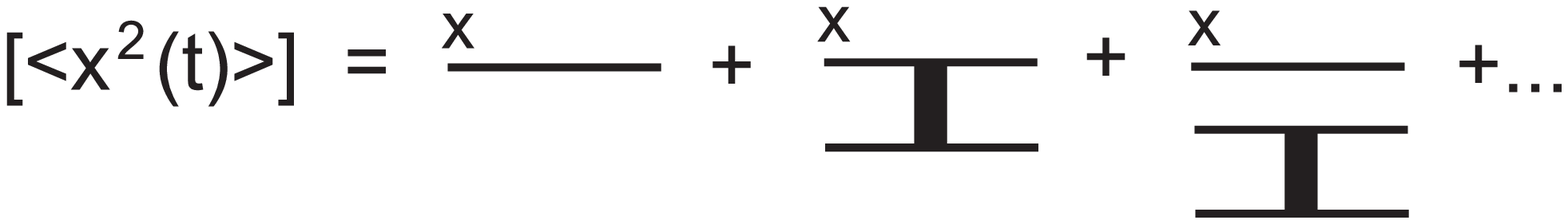}
\end{center}
\caption{Examples of skeleton graphs contributing to the transversal
displacement.}
\label{pdm_x^2_reg}
\end{figure}

\begin{figure}[tbph]
\begin{center}
\includegraphics[scale=0.4]{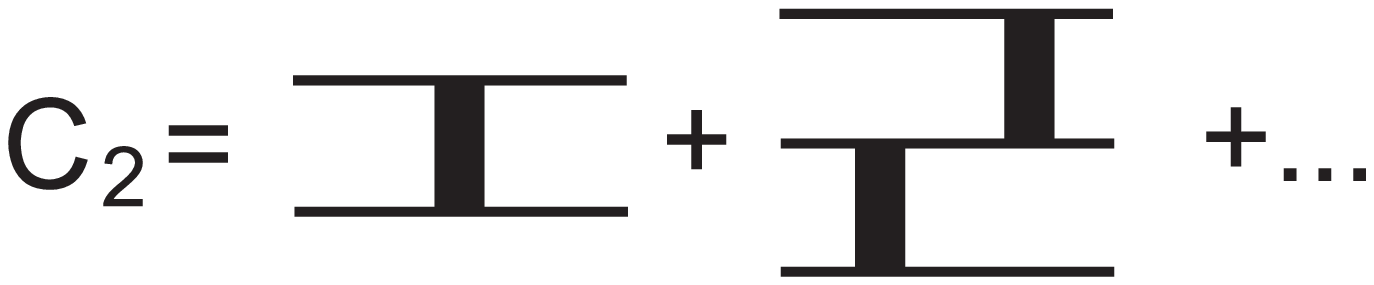}
\end{center}
\caption{Examples of skeleton graphs contributing to the 2nd cumulant of the
free energy.}
\label{pdm-C2}
\end{figure}

We now will perform the partial summation of perturbation series, which take
into account insertions associated with the $\left[ Z^{2}\right] _{c}$%
-vertex shown in Fig. \ref{pdm_Z^2_c}, and results in skeleton graphs, where
the bare interaction is replaced by the effective coupling constant $\Delta
(t_{2}-t_{1})$. The possibility for such a resummation is based on the
fact that the numerical prefactors of the graphs under resummation are the
same. Figs. \ref{pdm_x^2_reg}-\ref{pdm-C2} give examples of skeleton graphs
contributing to the mean-square displacement and the cumulants $C_{2}$.

The expressions for the mean-square displacement and the cumulant $C_{2}$
associated with the graphs in Figs. \ref{pdm_x^2_reg}-\ref{pdm-C2} are given
by
\begin{equation}
\left[ <x^{2}(t)>\right] =d\gamma t+d\gamma t\Gamma _{1}-\Gamma _{1,x}+\cdots,
\label{x^2t}
\end{equation}%
\begin{equation}
C_{2}=\Gamma _{1}-2\Gamma _{2}+\frac{3}{2}\Gamma _{1}^{2}+\cdots.  \label{C-2}
\end{equation}%
The graphs in Figs. \ref{pdm_x^2_reg}-\ref{pdm-C2} and in Eqs.~(\ref{x^2t}-%
\ref{C-2}) appear in the same order from left to right. The identification
of graphs with analytical expressions occurs according to existing rules.
The effective vertex in these graphs is associated with the expression $G_{0}(%
\sqrt{2}(x_{2}-x_{1}),t_{2}-t_{1})\Delta (t_{2}-t_{1}).$ As an example we
give the analytical expression associated with $\Gamma _{1}$
\begin{eqnarray}
&&\Gamma _{1}=\int_{0}^{t}dt_{2}\int_{0}^{t_{2}}dt_{1}\int d^{d}x_{a}\int
d^{d}x_{b}\int d^{d}x_{2}\int d^{d}x_{1}  \notag \\
&&G_{0}(x_{a},t-t_{2})G_{0}(x_{b},t-t_{2})G_{0}(\sqrt{2}\left(
x_{2}-x_{1}\right) ,t_{2}-t_{1})  \notag \\
&&\Delta (t_{2}-t_{1})G_{0}^{2}(x_{1},0,t_{1}).  \label{dp9}
\end{eqnarray}%
Carrying out the integrations over space variables results in%
\begin{equation}
\Gamma _{1}=2^{-d}\int_{0}^{t}dt_{2}\int_{0}^{t_{2}}dt_{1}\Delta
(t_{2}-t_{1})(2\pi \gamma t_{1})^{-d/2}.  \label{dp10}
\end{equation}%
Similarly, the analytical expression associated with $\Gamma _{1,x}$ is
obtained as%
\begin{eqnarray}
\Gamma _{1,x} &=&d\gamma
2^{-d}\int_{0}^{t}dt_{2}\int_{0}^{t_{2}}dt_{1}\left( t-t_{2}+\frac{%
t_{2}-t_{1}}{2}+\frac{t_{1}}{4}\right)   \notag \\
&&\Delta (t_{2}-t_{1})(2\pi \gamma t_{1})^{-d/2}  \label{dp11}
\end{eqnarray}%
and consequently for the quantity
\begin{eqnarray}
-\Gamma _{1,x}+d\gamma t\Gamma _{1} &=&d\gamma
2^{-d}\int_{0}^{t}dt_{2}\int_{0}^{t_{2}}dt_{1}\left( \frac{t_{2}-t_{1}}{2}+%
\frac{3}{4}t_{1}\right)   \notag \\
&&\Delta (t_{2}-t_{1})(2\pi \gamma t_{1})^{-d/2},  \label{dp12}
\end{eqnarray}%
which yields the first-order correction to the mean-square displacement of
the free end of DP in the skeleton expansion.

The computation of $\Gamma _{1}$, $d\gamma t\Gamma _{1}$, $-\Gamma _{1,x}$
and $C_{2}$ in $d=1$ for large $t$ results in%
\begin{equation}
\Gamma _{1}\simeq \sqrt{2}e^{t\text{$\Delta $}_{0}^{2}/4\gamma }-\frac{1}{%
\sqrt{2}}-\frac{1}{\text{$\Delta $}_{0}}\sqrt{\frac{2}{\pi }}\left( \frac{%
\gamma }{t}\right) ^{1/2}+\cdots,  \label{dp13}
\end{equation}%
\begin{eqnarray}
\gamma t\Gamma _{1},-\Gamma _{1,x} &\simeq &\frac{\gamma t}{\sqrt{2}}e^{t%
\text{$\Delta $}_{0}^{2}/4\gamma }-\frac{9\gamma ^{2}}{2\sqrt{2}\text{$%
\Delta $}_{0}^{2}}e^{t\text{$\Delta $}_{0}^{2}/4\gamma }+  \notag \\
&&\frac{t\gamma }{16\sqrt{2}}+\frac{5\gamma ^{3/2}}{4\sqrt{2\pi }\text{$%
\Delta $}_{0}}\sqrt{t}+\cdots,  \label{dp14}
\end{eqnarray}%
and%
\begin{equation}
C_{2}\simeq \frac{16}{\sqrt{3\pi }}\frac{\sqrt{\gamma t}}{\text{$\Delta $}%
_{0}}e^{t\text{$\Delta $}_{0}^{2}/4\gamma }-\left( \frac{56}{9}+\frac{8}{%
\sqrt{3}\pi }\right) \frac{\gamma }{\text{$\Delta $}_{0}^{2}}e^{t\text{$%
\Delta $}_{0}^{2}/4\gamma }+\frac{2\gamma }{\text{$\Delta $}_{0}^{2}}+\cdots
\label{dp15}
\end{equation}%
The computation of higher cumulants is similar, but is more time-consuming.
The skeleton perturbation series permit to compute the quantities under
consideration ($\left[ <x^{2}(t)>\right] $, $C_{2}$, $C_{3}$, $\cdots$) in the
strong coupling regime at arbitrary dimensions. The skeleton perturbation
expansions possess, however, the deficiency which is due to the exponential
dependence of the effective coupling constant on $t$. For sufficiently small
time $t$ the corrections are, however, small. While the connected graphs
contain the factor $\exp (p_{c}t)$, the disconnected ones the factor $\exp
(kp_{c}t)$, where $k$ is the number of connected parts of the disconnected
graph. The expressions (\ref{dp13}-\ref{dp15}) show that besides the
exponential terms the expansion in the skeleton series occurs in inverse
powers of $\Delta _{0}$. Just above the transition (in $d=1$ it corresponds
to $\Delta _{0}$ $\simeq 0$) the exponential terms are small, but the
corrections to them are correspondingly large.

We now will discuss the question how to proceed with the exponential terms, $%
\exp (p_{c}t)$, in skeleton expansions. A similar but however much simpler
problem is the localization of a directed polymer in an external $\delta $%
-potential $U(x)=-u\delta (x)$ studied using the perturbation expansion in
powers of $u$. The perturbation series of $<x^{2}(t)>$ is given in Fig. \ref%
{pdm_expot}.

\begin{figure}[tbph]
\begin{center}
\includegraphics[scale=0.5]{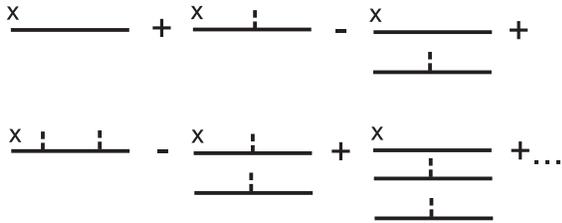}
\end{center}
\caption{
Perturbation series for DP in external potential.
}
\label{pdm_expot}
\end{figure}
A straightforward summation of the series results in the exact expression of the
mean-square displacement $<x^{2}(t)>=Z(t)/N(t),$ where the Laplace
transforms of the numerator and denominator are respectively given by
\begin{eqnarray*}
Z(p) &=&\frac{2d\gamma }{p^{2}}(1-uG_{0}(0,p;0))^{-1}, \\
N(p) &=&\frac{1}{p}(1-uG_{0}(0,p;0))^{-1}.
\end{eqnarray*}%
The exponential terms in the perturbation expansion cancel in the exact
expression. The DPRM can be considered qualitatively as a localization on
the randomly distributed $\delta $-potentials, which suggests that an
additional partial summations of skeleton expansion might be necessary to
eliminate the exponential terms, $\exp (kp_{c}t)$.
However, the
available terms in the skeleton expansions (\ref{dp13}-\ref{dp15}) make it
difficult to recognize the mechanism according to which such a resummation
has to be carried out. Next-order terms in the skeleton expansions would
simplify establishing the mechanism for performing such a resummation. While
the bare perturbation expansions in powers of $\Delta _{0}$ are not
alternating series, the skeleton series (\ref{dp13}-\ref{dp15}) are alternating ones. This property of skeleton expansions supports indirectly the idea
of a resummation.

To conclude, we have
performed partial summations of perturbation expansions of the directed polymer in disordered media to represent them as skeleton expansions in powers of the effective coupling constant, which corresponds to the binding state between two replicas or equivalently to the binding state of quantum particle in a delta potential. In the strong coupling phase the effective coupling constant exponentially increases with time. We argue that an additional resummation is necessary to cancel these exponential terms. The problem of elimination of the exponential terms from the skeleton expansions, which physically is expected to correspond to the redefinition of the ground state, represents a new paradigm and remains unsolved yet. We explicitly compute the mean-square displacement of polymer end and the second cumulant of the free energy to the lowest orders of the skeleton expansions.

\begin{acknowledgments}
A financial support from the Deutsche Forschungsgemeinschaft, the grant Ste
981/3-1, is gratefully acknowledged.
\end{acknowledgments}

\end{document}